\documentclass[%
reprint,
amsmath,amssymb,
aps,
longbibliography,
]{revtex4-1}

\usepackage{upgreek}
\usepackage{graphicx}
\usepackage{dcolumn}
\usepackage{bm}
\usepackage{color}
\usepackage[normalem]{ulem}
\usepackage{mathtools}
\usepackage[english]{babel}
\usepackage{xr-hyper}
\usepackage[colorlinks=true,linkcolor=blue,urlcolor=blue,citecolor=blue]{hyperref}

\let\Dref\relax
\let\mkmin\relax
\let\mkmax\relax
\let\fig\relax
\let\Ref\relax
\let\red\relax

\begin{document}

\title{Fragile Glasses Associated with a Dramatic Drop of Entropy under Supercooling}

\author{Chun-Shing Lee$^1$}
\author{Matteo Lulli$^{1,2}$}
\author{Ling-Han Zhang$^3$}
\author{Hai-Yao Deng$^4$}
\author{Chi-Hang Lam$^1$}
\email[Email: ]{C.H.Lam@polyu.edu.hk}
\address{$^1$Department of Applied Physics, Hong Kong Polytechnic University, Hong Kong, China \\
	$^2$Department of Mechanics and Aerospace Engineering, Southern University of Science and Technology, Shenzhen, Guangdong 518055, China \\
	$^3$Department of Physics, Carnegie Mellon University, Pittsburgh, Pennsylvania 15213, USA \\
	$^4$School of Physics and Astronomy, Cardiff University, 5 The Parade, Cardiff CF24 3AA, Wales, United Kingdom
}

\date{\today}

\newcommand{\blue}[1]{{\color{blue}{#1}}}
\newcommand\cmt[1]{\textcolor{red}{[#1]}}
\newcommand\move[1]{\textcolor{blue}{[#1]}}
\newcommand{\red}[1]{{\color{red}{#1}\color{black}}}
\newcommand{\chg}[1]{{\color{blue}{#1}}}        
\newcommand{\del}[1]{\sout{#1}}  
\newcommand{\delete}[1]{\sout{#1}}  
\newcommand{\rpl}[2]{{\sout{#1}}{\color{blue}{#2}}}        
\newcommand{\dis}[1]{{\color{red}\del{#1}}} 
\newcommand{\chk}[1]{{\color{red}\{? #1\}}}
\newcommand{\qn}[1]{{\color{red}\{#1 \}}}

\newcommand{\refb}[1]{(\ref{#1})}
\newcommand{\Ref}[1]{Ref.~\cite{#1}}
\newcommand{\fig}[1]{Fig.~\ref{#1}}
\newcommand{\figs}[2]{Figs.~\ref{#1} and \ref{#2}}
\newcommand{\Fig}[1]{Figure~\ref{#1}}
\newcommand{\eq}[1]{Eq.~(\ref{#1})}
\newcommand{\eqr}[2]{Eqs.~(\ref{#1})-(\ref{#2})}
\newcommand{\Eqr}[2]{Equations~(\ref{#1})-(\ref{#2})}
\newcommand{\eqs}[2]{Eqs.~(\ref{#1}) and (\ref{#2})}
\newcommand{\Eqs}[2]{Equations~(\ref{#1})-(\ref{#2})}
\newcommand{\Eq}[1]{Equation~(\ref{#1})}
\newcommand{\App}[1]{Appendix~\ref{#1}}
\renewcommand{\sec}[1]{Sec.~\ref{#1}}
\newcommand{\Sec}[1]{Section~\ref{#1}}

\newcommand{\roundbk}[1]{\left({#1}\right)}
\newcommand{\squarebk}[1]{\left[{#1}\right]}
\newcommand{\bracebk}[1]{\left\lbrace{#1}\right\rbrace}
\newcommand{\anglebk}[1]{\left\langle{#1}\right\rangle}
\newcommand{\lbracebk}[1]{\left\lbrace{#1}\right.}
\newcommand{\rbracebk}[1]{\left.{#1}\right\rbrace}

\newcommand{\phiv}{\phi_{v}}
\newcommand{\MSD}{\anglebk{|\mathbf{r}_{l}(t) - \mathbf{r}_{l}(0)|^{2}}}
\newcommand{\dE}{\Delta E}
\newcommand{\DV}{\Delta V}
\newcommand{\mkmin}{m_{k}^{strong}}
\newcommand{\mkmax}{m_{k}^{fragile}}

\newcommand{\Dref}{10^{-1}}

\renewcommand{\del}[1]{} 
\renewcommand{\red}[1]{#1}

\begin{abstract}
We perform kinetic Monte Carlo simulations of a distinguishable-particle lattice model of structural glasses with random particle interactions. 
By varying the interaction distribution and the average particle hopping energy barrier, we obtain an extraordinarily wide range of kinetic fragility. 
A stretching exponent, characterizing structural relaxation, is found to decrease with the kinetic fragility in agreement with experiments. 
The most fragile glasses are those exhibiting low hopping barriers and, more importantly, dramatic drops of entropies upon cooling toward the glass transition temperatures.
The entropy drops reduce possible kinetic pathways and lead to dramatic slowdowns in the dynamics.
In addition, the kinetic fragility is shown to correlate with a thermodynamic fragility.
\end{abstract}

\maketitle

\label{sec:Introduction}
An important concept in the study of structural glasses \cite{berthier2011review,garrahan2011review,stillinger2013review} is the kinetic fragility, often simply called the glass fragility, which has been investigated in great detail for different types of glass formers \cite{bohmer1993,angell1995,wang2006,giordano2003}.	
It describes how rapidly the dynamics slows down when temperature decreases.
The dynamics is typically characterized by viscosity, structural relaxation time \cite{angell1991,alba1990}, or particle diffusion coefficient \cite{fujara1992,coslovich2007}.
Glasses possessing the most dramatic slowdown are classified as fragile, whereas the opposite are referred to as strong.
Several models of glasses have been able to reproduce a range of kinetic fragilities \cite{garrahan2002geometrical,sausset2008,parmar2015,ozawa2016,ciarella2019}.
A closely related thermodynamic fragility \cite{angell2009} has also been defined and is based on how dramatically the entropy drops as the temperature decreases.
Experimental results indicate, in general, a positive correlation between the kinetic and thermodynamic fragilities \cite{angell2009,fontana2013}.
Yet, a fundamental understanding of the fragilities and their relationship is still lacking.

In this Letter, we study the kinetic and thermodynamic fragilities using a recently proposed distinguishable particle lattice model (DPLM) of structural glasses \cite{zhang2017}.
Lattice models are instrumental in statistical physics.
Celebrated examples include the Ising model for magnetism and the Edwards-Anderson model for spin glasses \cite{edwards1975}.
By bridging between analytic theory and more realistic models, they play pivotal roles in the solution and intuitive understanding of the systems concerned.
The DPLM aims at this bridging task for the study of structural glass.
It possesses exactly solvable equilibrium statistics \cite{zhang2017} and is promising for analytical treatment \cite{lam2018tree,deng2019}.
In support of its validity as a model of glass, DPLM has successfully reproduced typical glassy behaviors \cite{zhang2017}, a remarkable phenomenon known as Kovacs' expansion gap paradox \cite{lulli2020}, as well as Kovacs' effect for the aging of glasses \cite{lulli2019kovacseffect}. 
It captures in simpler and more tractable form the relevant physics seen in molecular dynamics (MD) and other realistic models, which in turn are more detailed  approximate models of glasses.
This should be a worthwhile approach considering that direct analytical treatment of MD or experimental systems in finite dimensions has proved exceedingly challenging and controversial \cite{stillinger2013review}.

Here, we show that both the kinetic and thermodynamic fragilities of the DPLM can be varied over wide ranges of values via the fine-tuning of its kinetic and thermodynamic properties.
Modeled glasses with higher kinetic fragilities in general exhibit smaller stretching exponents as well as higher thermodynamic fragilities, in good qualitative agreement with experiments.
The fundamental mechanisms behind the fragility variations in this model are intuitively understandable, and are likely applicable also to realistic glasses \cite{deng2019}.

\label{sec:The_model}
We adopt the DPLM proposed in Ref. \cite{zhang2017}, with minor differences explained in Sec.~I in supplementary information (SI).
It is defined on a 2D square lattice of size $L^{2}$ with $L=100$ and unit lattice constant following periodic boundary conditions.
There are $N$ distinguishable particles on the lattice labeled from 1 to $N$.
Each lattice site $i$ can be occupied by at most one of the particles with a particle index $s_{i} = 1,2,\dots,N$.
For unoccupied sites, i.e., sites occupied by voids, $s_{i} = 0$.
A void density of $\phi_v = 0.01$ is considered.
A particle configuration is specified by the set of particle indices $\bracebk{s_{i}}$ over all sites.
The total energy is
\begin{equation}
	E = \sum_{\anglebk{i,j}'} V_{s_{i}s_{j}},
	\label{E}
\end{equation}
where the sum is restricted to nearest neighbor (NN) sites $i$ and $j$ occupied by particles.
The interaction $V_{kl}$ for each pair of adjacent particles $k$ and $l$ is sampled before the start of the simulation from the pair-interaction distribution $g(V_{kl})$ and fixed subsequently.
The particle index $s_{i}$ is time dependent since the site $i$ will be visited by different particles as the system evolves.
Thus, $V_{s_{i}s_{j}}$ in \eq{E} is time dependent, although any $V_{kl}$ for any given particles $k$ and $l$ is quenched.
Dimensionless units will be adopted.

Particle distinguishability and particle-dependent interactions are directly justifiable for polydispersive or polymer systems. For identical-particle systems, it instead accounts effectively for the generally different frustration states experienced by the particles.
It also models high-entropy alloys \cite{yeh2013} in the limit of a large number of atomic species.
Being a lattice model, particle vibrations are not explicitly accounted for.
A particle configuration more precisely models an inherent state of a realistic system \cite{deng2019}.

\begin{figure}[t]
	\includegraphics[width=1\linewidth]{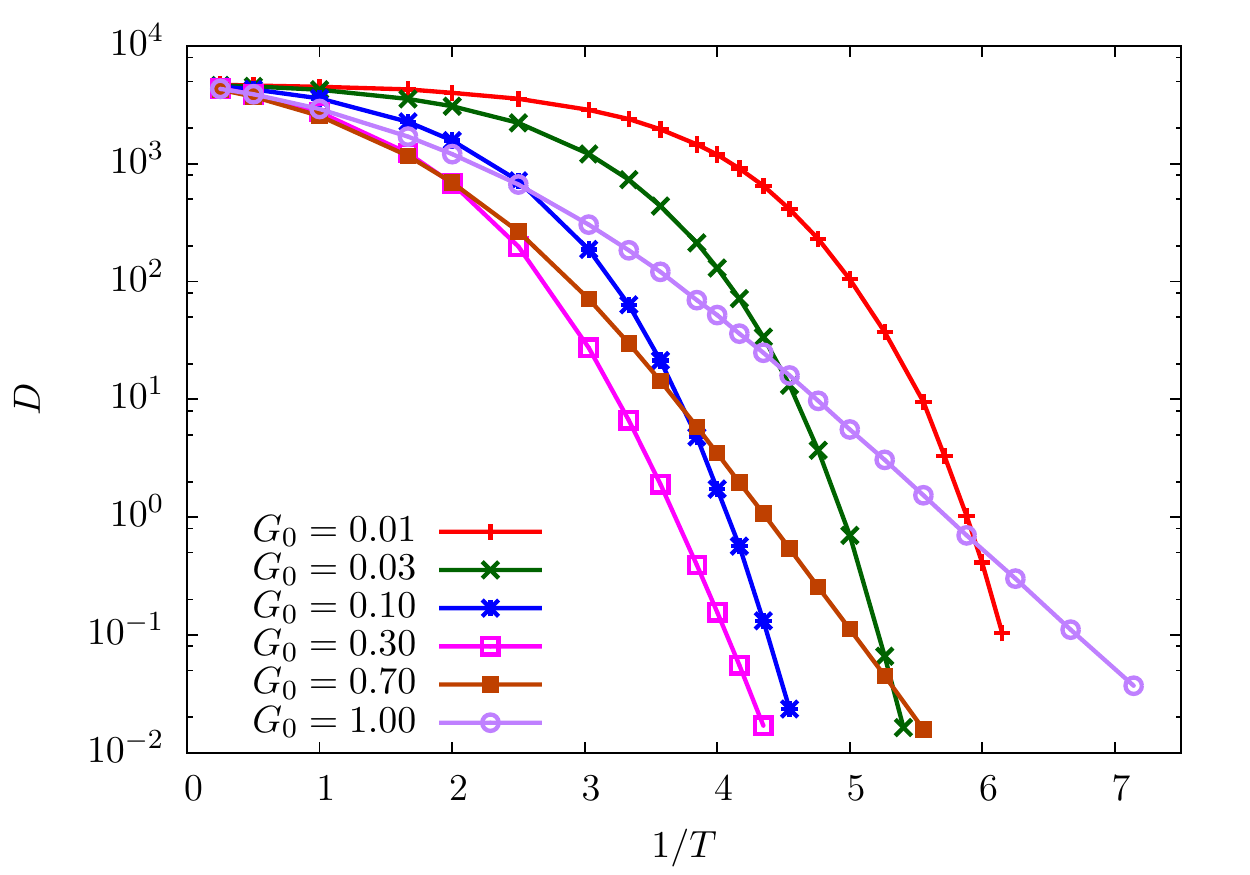}	
	\caption{Arrhenius plot of $D$ for various $G_{0}$ at $E_{0} = 0$. The system with a lower $G_{0}$ is more super-Arrhenius.}	
	\label{fig:DT}
\end{figure}

A main feature of our work is the random sampling for each $V_{kl} \in [V_0, V_1] \equiv [-0.5, 0.5]$ from
a \emph{bicomponent distribution} consisting of a uniform and a delta function representing, respectively, unexcited and excited states given by
\begin{equation}
	g(V) = \frac{G_{0}}{\Delta V} + (1 - G_{0})\delta(V - V_{1}),
	\label{gV}
\end{equation}
where $\Delta V = V_{1} - V_{0} =1$ and $\delta$ denotes the Dirac delta function.
\del{\eq{gV} is generalized from the uniform distribution adpoted in \cite{zhang2017}, where the newly added delta function is used for simplicity to represent a sharp peak at high interaction energy $V_1$.}
Here, $G_{0} \in [0,1]$ is our main thermodynamic parameter controlling the fragilities.
It equals the probabilistic weight of the uniform unexcited component of the distribution and also the probability density $g(V_0)$ at the ground state energy $V_0$.
For $G_{0}=1$, \eq{gV} reduces to the uniform distribution adopted in \Ref{zhang2017}, which leads to a strong glass.
Alternatively, for $G_{0}=0$, all interactions are at the excited energy state $V_1$ and the model reduces to a simple identical-particle lattice gas with a uniform particle interaction.
	


We assume a void-induced dynamics, which has been directly observed in recent experiments on glassy colloidal systems \cite{yip2020arxiv}. 
Using the Metropolis algorithm, each particle can hop to an unoccupied NN site at temperature $T$ at a rate
\begin{eqnarray}
	w = 
	\begin{cases}
		w_{0}\exp{\squarebk{-\roundbk{E_{0} + \dE}/k_BT}} & \text{for $\dE > 0$}, \\
		w_{0}\exp{\roundbk{-E_{0}/k_BT}}	& \text{for $\dE \leq 0$},
	\end{cases}
	\label{w}
\end{eqnarray}
where $\dE$ is the change in the system energy $E$ given by \eq{E} due to the hop and $k_B = 1$ is the Boltzmann constant.
We put $w_{0} = 10^{6}$.
The hopping energy barrier offset $E_{0} \geq 0$ is our main kinetic model parameter for controlling the fragilities.
Our algorithm satisfies detailed balance.

\begin{figure}[t]
	\includegraphics[width=1\linewidth]{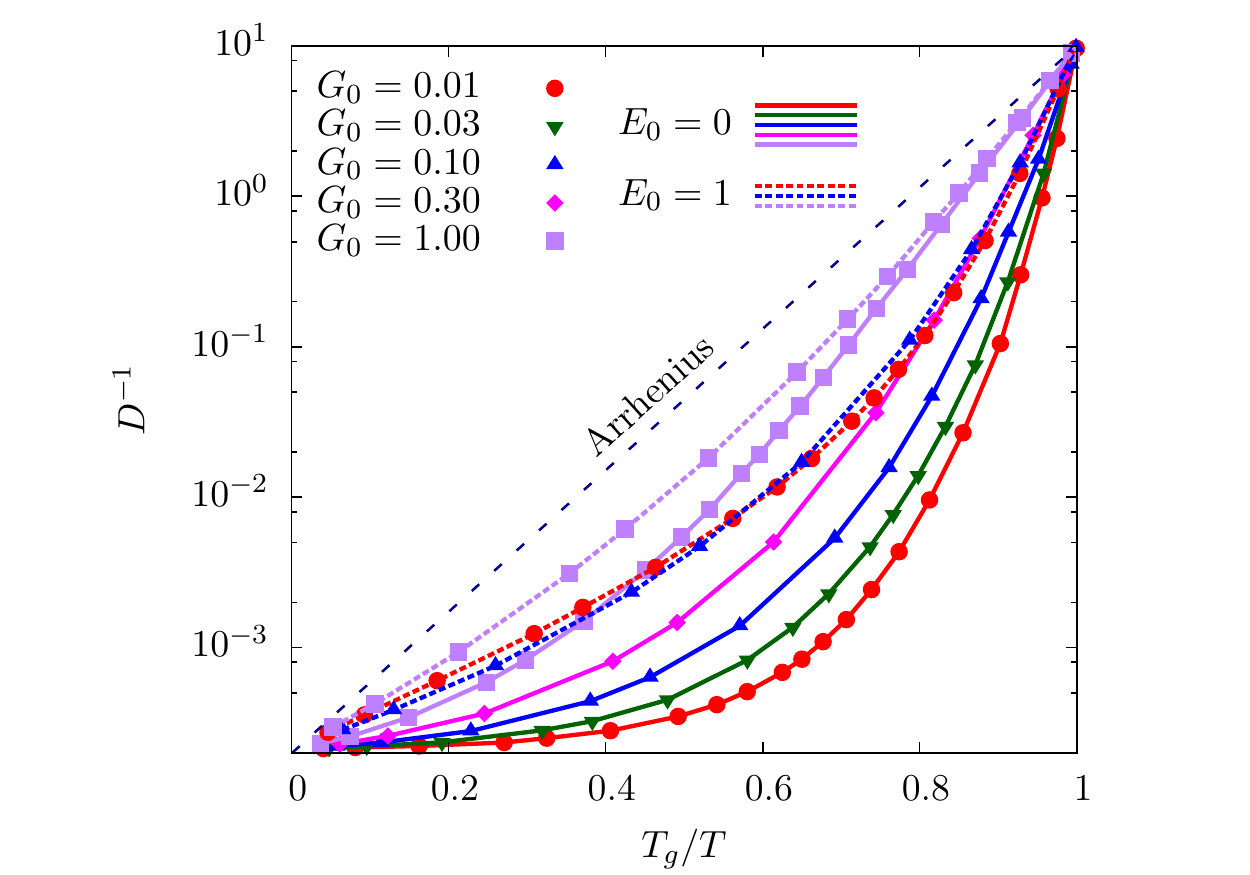}
	\caption{Kinetic Angell plot of $D^{-1}$ against $T_{g}/T$ for various $G_{0}$ and $E_0$, where $T_{g}$ for each curve is defined at $D_{r} = \Dref$. A low $G_{0}$ gives a fragile system. For a given $G_{0}$, increasing $E_{0}$ makes the system stronger.}
	\label{fig:AngellD}
\end{figure}

\label{sec:Results}
Kinetic Monte Carlo simulations have been performed on the DPLM, starting from directly constructed initial equilibrium configurations \cite{zhang2017}.
We report here our main results while further details are given in Sec.~II in SI.
The particle mean squared displacement defined as $\mbox{MSD}=\MSD$ is calculated, where $\mathbf{r}_{l}(t)$ denotes the position of particle $l$ at time $t$.
The particle diffusion coefficient $D$ is computed according to
$D = \roundbk{1/2d}\roundbk{\mbox{MSD}/t}$, where $d = 2$ is the dimension of the system, at sufficiently large values of $t$ in the diffusion regime. 

The Arrhenius plot in \fig{fig:DT} shows $D$ against $1/T$ for $E_{0} = 0$ and various $G_{0}$.
We observe that $\log D$ decreases with $1/T$ faster than linearly, demonstrating a super-Arrhenius slowdown.
The dependence of $D$ on $G_{0}$ for any given $T$ is nonmonotonic.
Yet, the super-Arrhenius behavior strengthens monotonically as $G_{0}$ decreases. 
This can be clearly seen in a \emph{kinetic} Angell plot in
\fig{fig:AngellD}, which plots $D^{-1}$ against $T_g/T$ for $E_0=0$ (solid lines) using the data from \fig{fig:DT}.
We have defined the glass transition temperature $T_g$ as $T$ at which $D= D_{r} \equiv \Dref$, where the reference diffusion coefficient $D_{r} $ is about the lowest value we can simulate.
We observe that $D$ now varies monotonically with $G_{0}$ for any given $T_g/T$. More importantly, the super-Arrhenius property clearly strengthens monotonically as $G_{0}$ decreases.
Related kinetic Angell plot of structural relaxation time extracted from  self-intermediate scattering function shows closely analogous trends (see Sec.~II in SI).

\Fig{fig:AngellD} also shows $D^{-1}$ for $E_0=1$ (dotted lines).
Results are simply obtained from values of $D$ for $E_0=0$ after rescaling time by a factor $\exp{(E_{0}/k_BT)}$, noting that $T_g$ has to be recalculated since $D_r$ is not rescaled.
We observe that a smaller $E_{0}$ strengthens the super-Arrhenius property at any given $G_{0}$. 
The results in \fig{fig:AngellD} capture many qualitative features in experimental findings \cite{angell1995,wang2006,giordano2003}.

The kinetic fragility $m_k$ describes the super-Arrhenius property quantitatively and is  defined by $m_{k} = \partial \log D^{-1}/\partial(T_{g}/T)|_{T = T_{g}}$. 
We obtain a wide range of values of $m_k$ from $6.76$ to $26.35$.
These values are in general smaller than experimental ones typically in the range from $25$ to $150$ \cite{micoulaut2016}, but this is only due to a rather small $D_r$ adopted for defining $T_g$.
An extrapolation to $D_r = 10^{-14}$ is performed  so that nearly   $18$ orders of magnitude of $D$ are considered, similar to analyses of structural relaxation time and viscosity in experiments \cite{angell1991,alba1990}.
Then, $m_k$ ranges from 21.4 for large $G_0$ and $E_0$ and 120 for $G_0=0.01$ and $E_0=0$, consistent with the experimental range
 (see Sec.~III in SI).


\begin{figure}[t]
	\includegraphics[width=1\linewidth]{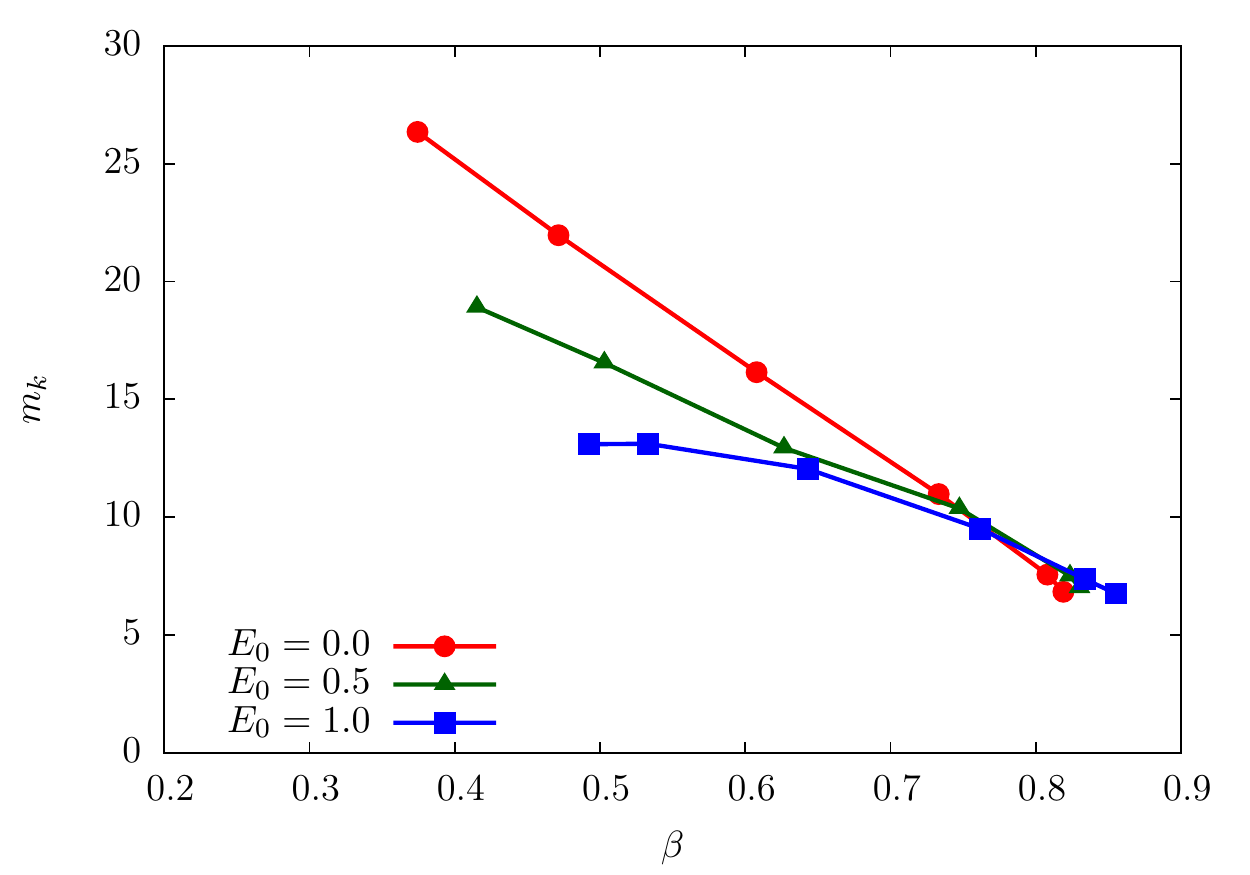}
	\caption{Relationship between $m_k$ and $\beta$ at $E_{0} = 0, 0.5, 1$ with $G_{0} = 0.01, 0.03, 0.1, 0.3, 0.7, 1$ (from left to right).}
	\label{fig:mkvsbetaTg}
\end{figure}

To further establish the physical relevance of the DPLM, we proceed to show that relaxation and thermodynamic properties of the strong and fragile glasses from this model are consistent with experiments.
First, structural relaxation is studied by measuring the self-intermediate scattering function 
\begin{equation}
	F_{s}(\mathbf{q},t) = \anglebk{e^{i\mathbf{q}\cdot(\mathbf{r}_{l}(t) - \mathbf{r}_{l}(0))}}, 
\end{equation}
where $q = (2\pi/L)q'$ with $q' = 10$.
The results are nicely fitted by the stretched exponential function $A\exp{\squarebk{-(t/\tau)^{\beta}}}$ for $t \agt \tau$, where $\beta$, $\tau$, and $A$ are, respectively, the stretching exponent, the relaxation time, and a constant close to unity.
\del{We plot $\beta$ against $T_{g}/T$ for various $G_{0}$ in \fig{S-fig:betaT}.}
\Fig{fig:mkvsbetaTg} plots $m_k$ against $\beta$ at $T_{g}$ for various $G_{0}$ and $E_0$. It shows that $m_k$ tends to decrease approximately linearly with $\beta$, in agreement with a trend observed previously in experiments \cite{micoulaut2016}.
In addition, the obtained range $0.37$ to $0.81$ of $\beta$ is comparable to that from experiments.
Results on $\beta$ are not significantly affected by using smaller values of $D_{r}$, especially for the fragile glasses since $T_{g}$ only changes slightly.

\begin{figure}[t]
	\includegraphics[width=1\linewidth]{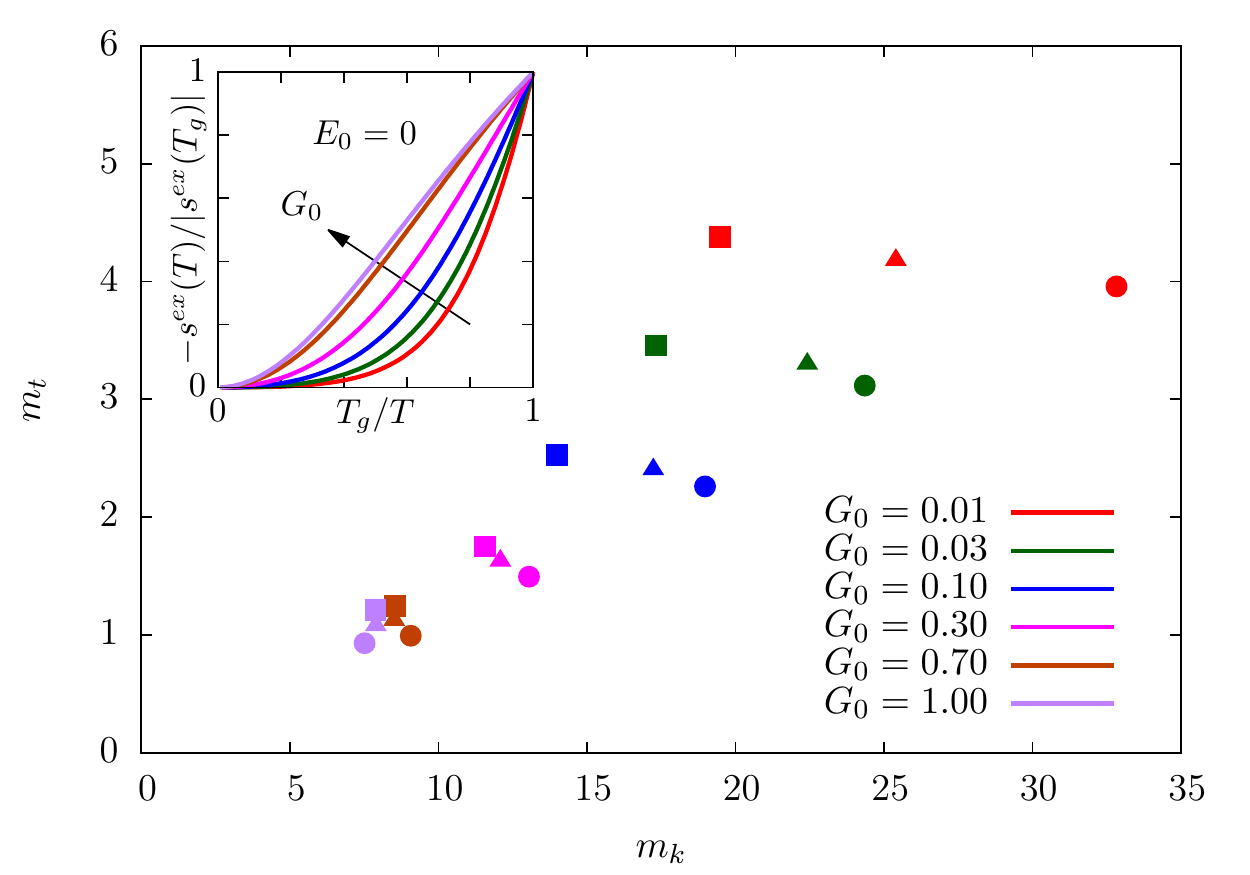}
	\caption{Plot of $m_{t}$ against $m_{k}$ at $E_{0} = 0$ (circle), $0.5$ (triangle), and $1$ (square) for various $G_{0}$. Inset: Thermodynamic Angell plot of $-s^{\text{ex}}(T)/|s^{\text{ex}}(T_{g})|$ at $E_{0} = 0$.}
	\label{fig:mtvsmk}
\end{figure}

Second, we study the thermodynamic properties of our model by calculating an entropy-based \emph{thermodynamic fragility}. 
The equilibrium statistics including the partition function $Z$ of the DPLM are exactly known \cite{zhang2017}.
The entropy per particle $s(T)$ is computed accordingly.
We further define an excess entropy per particle $s^{\text{ex}}(T) = s(T) - s^{\text{LG}}$ over the entropy $s^{\text{LG}}$ of a simple lattice gas \cite{dyre2018}. 
[See Eq.~(S23).]
The inset of \fig{fig:mtvsmk} shows a \emph{thermodynamic} Angell plot of $-s^{\text{ex}}(T)/|s^{\text{ex}}(T_{g})|$ against $T_{g}/T$ for $E_{0} = 0$ and different $G_{0}$.
The results resemble those of closely related thermodynamic Angell plots from experiments \cite{martinez2001} as well as the kinetic Angell plot in \fig{fig:AngellD}.
An increased $E_{0}$ alters the curvature only slightly for all values of $G_{0}$. 
In general, a strong glass with $G_{0}=1$ is also thermodynamically strong with a close-to-linear relation, while a fragile glass at $G_{0}=0.01$ shows the most dramatic variations.
The trend is in general similar if other forms of thermodynamic Angell plots \cite{martinez2001} are considered.

We define a thermodynamic fragility $m_{t}$ as 
\begin{equation}
	m_t = \left.\frac{\partial\roundbk{-s^{\text{ex}}(T)/|s^{\text{ex}}(T_{g})|}}{\partial\roundbk{T_{g}/T}} \right\rvert_{T = T_{g}},
	\label{mt}
\end{equation}
which is analogous to the kinetic counterpart $m_k$.
\del{A plot of $m_t$ against $G_{0}$ for various values of $E_0$ is shown in \fig{S-fig:mtvsg0overE0}.}
\Fig{fig:mtvsmk} shows the kinetic fragility $m_t$ against the thermodynamic fragility $m_k$ for various $G_{0}$ and $E_{0}$, displaying a clear tendency of a positive correlation, i.e., $m_t \sim m_k$.
The correlation is consistent with the general trend observed in experiments based on related definitions \cite{martinez2001} and is mainly caused by the similar dependencies of $m_k$ and $m_t$ on $G_{0}$.

\label{sec:Discussions}
We have studied glass fragility using the DPLM for various values of model parameters $G_{0}$ and $E_0$.
The most fragile glass is obtained at small $G_{0}$ and $E_0=0$.
Extrapolating our simulation results toward $G_{0} \to 0$, the kinetic fragility $m_k$ appears to rise unboundedly (see Fig.~S12).
The DPLM may hence model in-principle arbitrarily fragile glasses. 
Simulations at very small $G_{0}$ are, however, prohibitively intensive due to increased finite-size effects.
At $G_{0}=0$, the model reduces to a simple lattice gas, which is not glassy.
A high $m_k$ thus requires a small but nonvanishing probability of low-energy particle pairings.

\label{cmtA5B2}
We argue that $G_0$ is the main material parameter which captures the relevant particle interaction characteristics and determines the fragilities $m_t$ and $m_k$ in glasses.
The value of $G_0$ in a glass depends on the detailed molecular interactions and is strongly affected, for example, by the geometries of any tightly bounded groups of atoms.
A fragile glass obtained at a small $G_0$ can be intuitively understood as follows.
At high $T$, all particle configurations are possible, leading to a high entropy $s(T)$ independent of $G_0$. Most interactions take the excited states with energy $V_1$ due to their high probabilistic weight $1-G_0$ [see \eq{gV}].
Particle pairings with unexcited energies close to $V_0$ are in contrast rare due to the small probabilistic weight $G_0$. As $T$ decreases, the lower energies render them  energetically favorable and increasingly dominant.
The entropy $s(T)$ thus drops dramatically and becomes small at low $T$, accounting for a high $m_t$.

We further suggest that this high $m_t$ is closely correlated to a high $m_k$.
This is because the system dynamics at low $T$ amounts to sampling various energetically favorable configurations.
The rarity of these configurations as indicated by the low entropy implies highly constrained kinetic pathways of particle motions.
This leads to a sharp drop in $D$ as described by a large $m_k$.
A possible characteristic of constrained kinetics is repetitive particle motions.
Our picture is thus supported by a sharp increase in a particle return probability but a mild drop of the particle hopping rate as $T$ decreases as explained in Sec.~II in SI. 
The thermodynamic parameter $G_0$ therefore strongly impacts the system thermodynamics and hence also the kinetics. 
In contrast, the kinetic parameter $E_0$ is of lesser importance to the fragility properties.
By controlling the hopping barrier, it clearly has a strong and direct impact on $m_k$.
However, it plays no role in the equilibrium statistics and in particular in the system entropy [see Eq.~(S23)].
It has a tiny impact on $m_t$ only by influencing the value of $T_g$ at which $m_t$ is evaluated.
Not accounting for the correlation between $m_k$ and $m_t$ observed in experiments,
 we expect $E_0$ to play a smaller role in the variation of $m_k$ among various glasses.

The particle interaction distribution $g(V)$ has been taken with a bicomponent form consisting of a low-energy uniform distribution and a high-energy delta function for simplicity. 
The delta function represents excited particle interactions more relevant at higher $T$  and replacing it by some narrow Gaussian leads to similar simulation results. 
The uniform distribution is the simplest continuous distribution with a lower bound $V_0$, corresponding to the energy minimum present in typical pair potentials such as the Lennard-Jones potential.
The continuous form of $g(V)$ around $V_0$ is expected to lead to glassy behaviors even at a very low $T$, as the model reduces to one with a single uniform distribution studied in \Ref{zhang2017}.

The DPLM with a bicomponent $g(V)$ is closely related to a bond excitation model proposed by Moynihan and Angell \cite{moynihan2000}, in which particle bonds can assume either an unexcited or excited state {(see Sec.~VI in SI)}.
At low $T$, the realized interactions $V_{s_{i}s_{j}}$ from the uniform unexcited component have a small energy spread of about $k_BT$ around $V_0+k_BT$. Neglecting this energy spread, the ratio of the degeneracy of
the excited states to that of the unexcited states is about $(1-G_{0})/G_{0}$, leading to an entropy difference  
\begin{equation}
  \label{DS}
  \Delta S^0 \simeq k_B \ln[(1-G_{0})/G_{0}].
\end{equation}
Considering $G_{0}=0.01$ corresponding to fragile glasses, we get $\Delta S^0 \simeq 4.60 k_B $.
A more accurate calculation using Eq.~(S45) gives a similar value of $\Delta S^0 \simeq 5.42k_B$.
Reverting to physical units with $k_B=8.315$ J/mol K, it gives $\Delta S^0 \simeq 45.1$ J/K per mole of excitable states.
This value matches that of $\Delta S^0$, for example, for toluene in \Ref{moynihan2000}, which has a high $m_k=103$.
In addition, $\Delta H^0 \simeq 1-k_BT_g$ is the energy difference between the excited and unexcited states in our model. At $T_g \simeq 0.163$, $\Delta H^0/k_B T_g \simeq (1-0.163) / 0.163 \simeq 5.15$ for $G_{0}=0.01$.
It compares well with the value $6.95$ for toluene in \Ref{moynihan2000}.

The quantitative consistency demonstrated above means that the bond excitation model provides a simplified theoretical description for the thermodynamic properties of the DPLM with the bicomponent $g(V)$.
Moreover, the success of the bond excitation model in describing the entropy of fragile glasses in \Ref{moynihan2000}  justifies the bicomponent form of $g(V)$ used in this work.
From \eq{DS}, a fragile glass characterized by a small $G_0$ possesses a large $\Delta S^0$. These material parameters depend on the detailed molecular interactions. For molecular or polymer glasses which are often fragile, their values may reflect that the geometrically complex molecules fit well with each other to form very stable bonds only at a rare set of orientations and conformations. In contrast, strong glasses including network glasses may consist of simpler structures such as tetrahedrons. A simple random spread of the interactions due to frustration can then account for $G_0\simeq 1$ and a small $\Delta S^0$.

We have found that the thermodynamic parameter $G_{0}$ has the strongest impacts on both $m_k$ and $m_t$.
In contrast, the kinetic parameter $E_0$ also plays a significant role for $m_k$ but not so much for $m_t$.
Further simulations show that the void density $\phiv$ has rather small effects on both $m_k$ and $m_t$, as long as $\phiv \ll 1$ which ensures the glassy state.
One can also consider model variations such as a different $g(V)$.
Since glass properties depend on multiple model parameters, the relations discussed here between $m_k$, $m_t$, and $\beta$ are only general trends assuming small variations in other parameters.
Exceptions are thus possible in more general settings.
From another point of view, the value of $m_k$ does not uniquely determine the precise geometry of the whole curve in the Angell plot in \fig{fig:AngellD} when multiple material parameters are taken into account.
These are fully consistent with experimental observations \cite{angell1995}.  

To sum up, we have studied fragility properties of glasses using kinetic Monte Carlo simulations and analytic calculations based on the DPLM.
A wide range of values of kinetic fragility is reproduced, indicating the possibility of arbitrarily fragile glasses limited only by computational resources. 
The kinetic fragility is mainly controlled via a thermodynamic parameter $G_{0}$, dictating the probability distribution of particle pair interactions.
The most fragile glass is obtained at small $G_{0}$ corresponding to the case that pair interactions can take low-energy states with a small but nonvanishing probability, i.e. low-entropy unexcited states.
These configurations physically represent rare pairings between particles with exceptionally stable arrangements.
As the temperature decreases, particle configurations are increasingly constrained to these low-energy pairings.
This causes a dramatic drop in the entropy associated with a dramatic slowdown in the dynamics, resulting, respectively, in high thermodynamic and kinetic fragilities.
Our model, upon variations in $G_{0}$, exhibits correlations between kinetic fragility, thermodynamic fragility, and a relaxation stretching exponent, in qualitative agreement with general trends observed in experiments.
The kinetic fragility is also affected by a kinetic model parameter $E_0$.
A fragile glass is obtained at small $E_0$ corresponding to particle hopping activation barriers with an average which is small compared to their fluctuations.

We thank the support of Hong Kong GRF (Grant No. 15330516), Hong Kong PolyU (Grant No. 1-ZVGH), and National Natural Science Foundation of China (Grant No. 11974297).

~\\~

%

\end{document}